\documentclass[preprint2]{aastex}
\usepackage{graphicx}
\usepackage{epsfig}

\shorttitle{Sgr A$^*$: a laboratory to measure the central black
hole and cluster parameters} \shortauthors{De Paolis et al.}

\begin{document}


\title{Sgr A$^*$: a laboratory to measure the central black hole and
stellar cluster parameters}


\author{A.A. Nucita, F. De Paolis and G. Ingrosso}
\affil{Dipartimento di Fisica, Universit\`a di Lecce, and {\it
INFN}, Sezione di Lecce, CP 193, I-73100 Lecce, Italy}

\author{A. Qadir}
\affil{Center for Advanced Mathematics and Physics, National
University of Science and Technology, Campus of College of
Electrical and Mechanical Engineering, Peshawar Road, Rawalpindi,
Pakistan}

\and

\author{A.F. Zakharov\altaffilmark{1,2}}
\affil{Institute of Theoretical and Experimental Physics, 25,
B.Cheremushkinskaya St., Moscow, 117259, Russia}


\altaffiltext{1}{Astro Space Centre of Lebedev Physics Institute,
Moscow, Russia} \altaffiltext{2}{Bogoliubov Laboratory for
Theoretical Physics, Joint Institute for Nuclear Research, Dubna,
Russia}


\begin{abstract}
Several stars orbit around a black hole candidate of mass
$3.7\times 10^6$ M$_{\odot}$, in the region of the Galactic Center
(GC). Looking for General Relativistic (GR) periastron shifts is
limited by the existence of a stellar cluster around the black
hole that would modify the orbits due to classical effects that
might mask the GR effect. Only if one knows the cluster parameters
(its mass and core radius) it is possible to unequivocally deduce
the GR effects expected and then test them. In this paper it is
shown that the observation of the proper motion of Sgr~A$^*$,
$v_{Sgr~A^*} = (0.4\pm 0.9)$ km s$^{-1}$ (\citealt{reid2004}),
could help us to constrain the cluster parameters significantly
and that future measurements of the periastron shifts for at least
three stars may adequately determine the cluster parameters and
the mass of the black hole.
\end{abstract}





\section{Introduction}

GR predicts that orbits about a massive central body suffer
periastron shifts yielding {\textit {rosette}} shapes. However,
the classical perturbing effects of other objects on inner orbits
give an opposite shift. Since the periastron advance depends
strongly on the compactness of the central body, the detection of
such an effect may give information about the nature of the
central body itself. This would apply for stars orbiting close to
the GC, where there is a \lq\lq dark object", the black hole
hypothesis being the most natural explanation of the observational
data. A cluster of stars in the vicinity of the GC (at a distance
$< 1\arcsec $) has been monitored by ESO and Keck teams for
several years (\cite{Genzel03,Schoedel03,Ghez03,Ghez04,Ghez05}).
In particular, Ghez et al. (\citeyear{Ghez03}) have reported on
observations of several stars orbiting close to the GC massive
black hole. Among those, the S2 star, with mass $M_{\rm S2}\simeq
15$ M$_{\odot}$, appears to be a main sequence star orbiting the
black hole with a Keplerian period of $\simeq 15$ yrs. This yields
\citet{Ghez05} a mass estimate of $M_{\rm Sgr~A^*}\simeq
3.67\times 10^6 ~M_{\odot}$ within $4.87\times 10^{-3}$ pc, that
is the S2 semi-major axis.

Several authors have discussed the possibility  of measuring the
GR corrections to Newtonian orbits for Sgr~A$^*$ (see e.g.
\citealt{Jaroszynski98, Jaroszynski99, Jaroszynski00, Fragile00,
Rubilar01, Weinberg05}), usually assuming that the central body is
a Schwarzschild black hole. However, since black holes generally
rotate, and there is no reason why they should not be rotating
fast, the Kerr metric should be used instead. Not only stellar
mass black holes but also supermassive black holes are believed to
be spinning. Indeed, X-ray observations of Seyfert galaxies,
microquasars and binary systems (\cite{fabian1, tanaka1, Fabi00,
Fabian04} and references therein) show that the data could be
explained by a rotating black hole model (see e.g.
\cite{zak_rep03_aa, Zak_rep03_AR,ZKLR02} and \cite{ ZR_ASR04}).
Further, supermassive black holes at the center of QSOs, AGNs and
galaxies show beamed jet emission implying that they have non zero
angular momentum. Hence, Kerr black holes may be fairly common in
nature. The relatively short orbital period of the S2 star
encourages a search for genuine GR effects like the orbital
periastron shift. Quite possibly, more suitable stars, close to
the GC black hole, will be found in the future. Bini et al.
(\citeyear{bini2005}) studied the GR periastron shift around
Sgr~A$^*$ and estimated it for various solutions belonging to the
Weyl class, including the Schwarzschild and Kerr black holes.
However, they did not take into account the presence of a stellar
cluster, which could in principle be sizable.

The purpose of this paper is to try to find limits for the extent
and density of the cluster about Sgr~A$^*$ and if those limits
yield a measurement of its spin.

Clearly, a thorough knowledge of the cluster mass and density
distribution is necessary to be able to infer the mass and spin of
the black hole at the GC by measuring the periastron shift and
subtracting the Newtonian shift. Unfortunately, the star cluster
parameters are poorly known. \footnote{We remark that the star
cluster we are considering around the central black hole might
contain not only normal stars but also white dwarfs, neutron stars
and/or stellar mass black holes.} However, the measure of the
Brownian motion of the central black hole due to the surrounding
matter may be used to constrain the black hole to cluster mass
ratio.\footnote{Other methods for estimating the black hole
parameters (i.e. mass and angular momentum) based on gravitational
retrolensing have been proposed. For more details on this topic we
refer to \citet{depaolis1,depaolis2,depaolis3,depaolis4,ZNDI_04}
and reference therein.} The latest observations of the $Sgr~A^*$
proper motion, $v_{Sgr~A^*}= (0.4\pm 0.9)$ km $\rm s^{-1}$
(\citealt{reid2004}), is much tighter than the earlier one of
$2\mathrm{-}20$ km $\rm{s}^{-1}$ (see Reid, Readhead, Vermeulen et
al. \citeyear{reid1999}).

For a test particle orbiting a Schwarzschild black hole of mass
$M_{\rm BH}$, the periastron shift is given by (see e.g. Weinberg,
\citeyear{Weinberg72})
\begin{equation}
\Delta \phi_S \simeq \frac{6\pi G
M_{BH}}{d(1-e^2)c^2}+\frac{3(18+e^2)\pi
G^2M_{BH}^2}{2d^2(1-e^2)^2c^4}~, \label{schshift}
\end{equation}
$d$ and $e$ being the semi-major axis and eccentricity of the test
particle orbit, respectively. For a rotating black hole with spin
parameter $a=|{\bf a}|=J/GM_{\rm BH}$, the space-time is described
by the Kerr metric and, in the most favorable case of equatorial
plane motion ({\bf a.v} = 0), the shift is given by (Boyer and
Price \citeyear{boyerprice}, but see also Bini et al.
\citeyear{bini2005} for more details)
\begin{equation}
\begin{array}{l}
\displaystyle{\Delta \phi_K \simeq \Delta \phi_S +\frac{8a\pi
M_{BH}^{1/2}G^{3/2}}{d^{3/2}(1-e^2)^{3/2}c^3}+\frac{3a^2\pi
G^2}{d^{2}(1-e^2)^{2}c^4}~,} \label{kershift}
\end{array}
\end{equation}
which reduces to eq. (\ref{schshift}) for $a\rightarrow 0$. In the
more general case, {\bf a.v} $\neq 0$, the expected periastron
shift has to be evaluated numerically.

The expected periastron shifts (mas/revolution), $\Delta \phi$ (as
seen
 from the center) and $\Delta \phi _E$ (as seen from Earth at
the distance $R_0\simeq~8$ kpc from the GC), for the Schwarzschild
and the extreme Kerr black holes, for the S2 and S16 stars turn
out to be $\Delta\phi^{S2}=6.3329\times 10^5$ and $6.4410\times
10^5$ and $\Delta \phi _E^{S2}=0.661$ and $0.672$ respectively,
and $\Delta\phi^{S16}=1.6428\times 10^6$ and $1.6881\times 10^6$
and $\Delta \phi _E^{S16}=3.307$ and $3.399$ respectively. Recall
that
\begin{equation}
\Delta \phi _E = \frac{d(1+e)}{R_0} \Delta \phi_{S,K}~.
\end{equation}
Notice that the differences between the periastron shifts for the
Schwarzschild and the maximally rotating Kerr black hole is at
most $0.01$ mas for the S2 star and $0.009$ mas for the S16 star.
In order to make these measurements with the required accuracy,
one needs to know the S2 orbit with a precision of at least $10$
$\mu$as.

There is a proposal to improve the angular resolution of VLTI with
the PRIMA facility (\cite{Rottgering03, Delplancke03,
Quirrenbach03} but see also the related
web-site\footnote{http://obswww.unige.ch/PRIMA/home/introduction.}),
which, by using a phase referenced imaging technique, will get
$\sim 10$ $\mu$as angular resolution. Hence, at least in
principle, the effect of a maximally rotating Kerr black hole on
the periastron shift of the S2 star can be distinguished from that
produced by a Schwarzschild black hole with the same mass.

The plan of the paper is as follows: In the next section we
briefly discuss the effect of a central star cluster on the
periastron advance. In Section 3 we use the Sgr~A$^*$ Brownian
motion to constrain the black hole to star cluster mass ratio.
Then we consider whether the detection of the spin of the black
hole from the periastron shift of the S2 star is possible, once
the cluster density and size have been adequately constrained. In
Section 5 we show how future measurements of the periastron shifts
for at least three stars close to the GC black hole may be used to
estimate the black hole mass and the star cluster mass density
distribution. In the next section we consider what the
observational requirements would be for adequate determination of
the cluster parameters to be able to resolve the Kerr effect.
Finally, in section 7, we present some concluding remarks.

\section{Retrograde shift due to a central stellar cluster}

The star cluster surrounding the central black hole in the GC
could be sizable. At least 17 members have been observed within 15
mpc up to now (\citealt{Ghez05}). However, the cluster mass and
density distribution, that is to say its mass and core radius, is
still unknown. The presence of this cluster affects the periastron
shift of stars orbiting the central black hole. The periastron
advance depends strongly on the mass density profile and
especially on the central density and typical length scale.

We model the stellar cluster by a Plummer model density profile
(\citealt{binneytremaine})
\begin{equation}
\rho_{CL}(r) = \rho_0 f(r)~,~~~~~~~~{\rm
with}~~~~~~~~~~f(r)={\left[1+\left(\frac{r}{r_c}\right)^2\right]^{-\alpha/2}}~,\label{plummer1}
\end{equation}
where the cluster central density $\rho _0$ is given by
\begin{equation}
\rho _0 = \frac{M_{CL}}{\int _0^{R_{CL}} 4\pi r^2 f(r)~dr}~,
 \label{plummer2}
\end{equation}
$R_{CL}$ and $M_{CL}$ being the cluster radius and mass,
respectively. According to dynamical observations towards the GC,
we require that the total mass $M(r)=M_{BH}+M_{CL}(r)$ contained
within $r\simeq 5\times 10^{-3}$ pc is $M\simeq 3.67\times 10^
6~M_{\odot}$. Useful information is provided by the cluster mass
fraction, $\lambda_{CL}=M_{CL}/M$, and its complement,
$\lambda_{BH}=1-\lambda_{CL}$. As one can see, the requirement
given in eq. (\ref{plummer2}) implies that $M(r)\rightarrow
M_{BH}$ for $r\rightarrow 0$. The total mass density profile
$\rho(r)$ is given by
\begin{equation}
\rho(r) = \lambda_{BH} M \delta ^{(3)}(\overrightarrow{r}) +\rho_0
f(r)~ \label{totaldensity}
\end{equation}
and the mass contained within $r$ is
\begin{equation}
M(r) = \lambda_{BH}M + \int_0^r 4\pi r'^2\rho_0 f(r')~dr'~.
\end{equation}
\begin{figure}
\begin{center}
\includegraphics[scale=0.50]{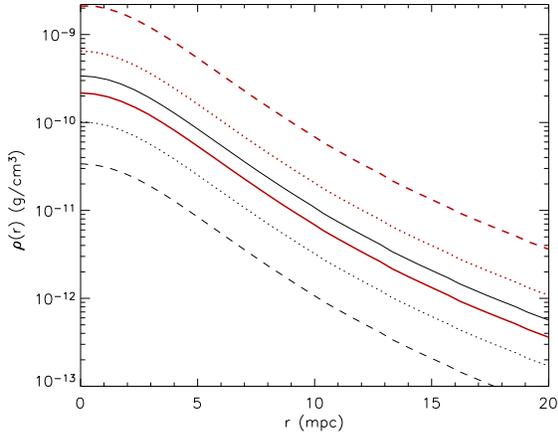}\qquad
\end{center}
\caption{The cluster mass density profile is shown for different
values of $\lambda_{BH}$. Solid, dotted and dashed lines
correspond to $\lambda_{BH}= 0,~0.7,~0.9$, respectively. Thick,
red lines have been obtained for $r_c=3$ mpc with the same values
$\lambda_{BH}$ as given above.} \label{density}
\end{figure}
\begin{figure}
\begin{center}
\includegraphics[scale=0.50]{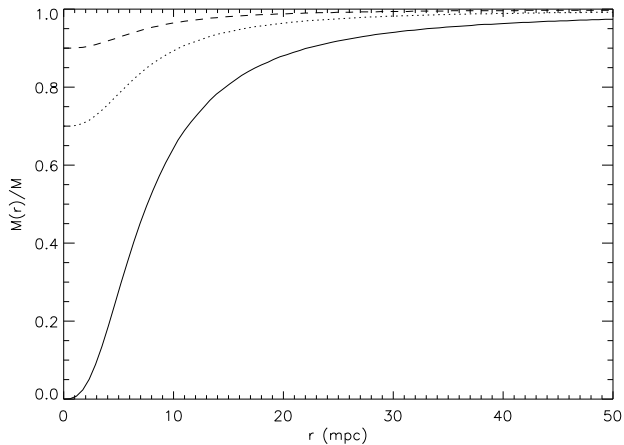}\qquad
\end{center}
\caption{The mass enclosed within the distance $r$ is shown for
different fractions $\lambda_{BH}$ of the total mass $M$ which is
contained in the central black hole. Solid, dotted and dashed
lines correspond to $\lambda_{BH}= 0$, $\lambda_{BH}= 0.7$ and
$\lambda_{BH}=0.9$, respectively.
Note that the case corresponding to $\lambda_{BH} = 0$ is not
realistic as shown by some observations (\cite{shen}).}
\label{ratiomass}
\end{figure}
In Figure \ref{density} we show the cluster mass density profile
$\rho_{CL}(r)$ as given by eq. (\ref{plummer1}), for selected
values of $\lambda_{BH}$. The total mass $M(r)$ enclosed within
the radius $r$ is also shown in Figure \ref{ratiomass}. In both
Figures, solid, dotted and dashed lines correspond to
$\lambda_{BH}= 0,~0.7,~0.9$, and we have assumed $r_c=3$ mpc
(thick lines) and $r_c=5.8$ mpc (thin lines).

The Newtonian gravitational potential $\Phi_N$ at a distance $r$
due to the mass contained within it can be evaluated as
\begin{equation}
\Phi_N(r) = -G \int_r^{\infty} \frac{M(r')}{r'^2}~dr'~.
\label{gravitationalpotential}
\end{equation}
\begin{figure}
\begin{center}
\includegraphics[scale=0.50]{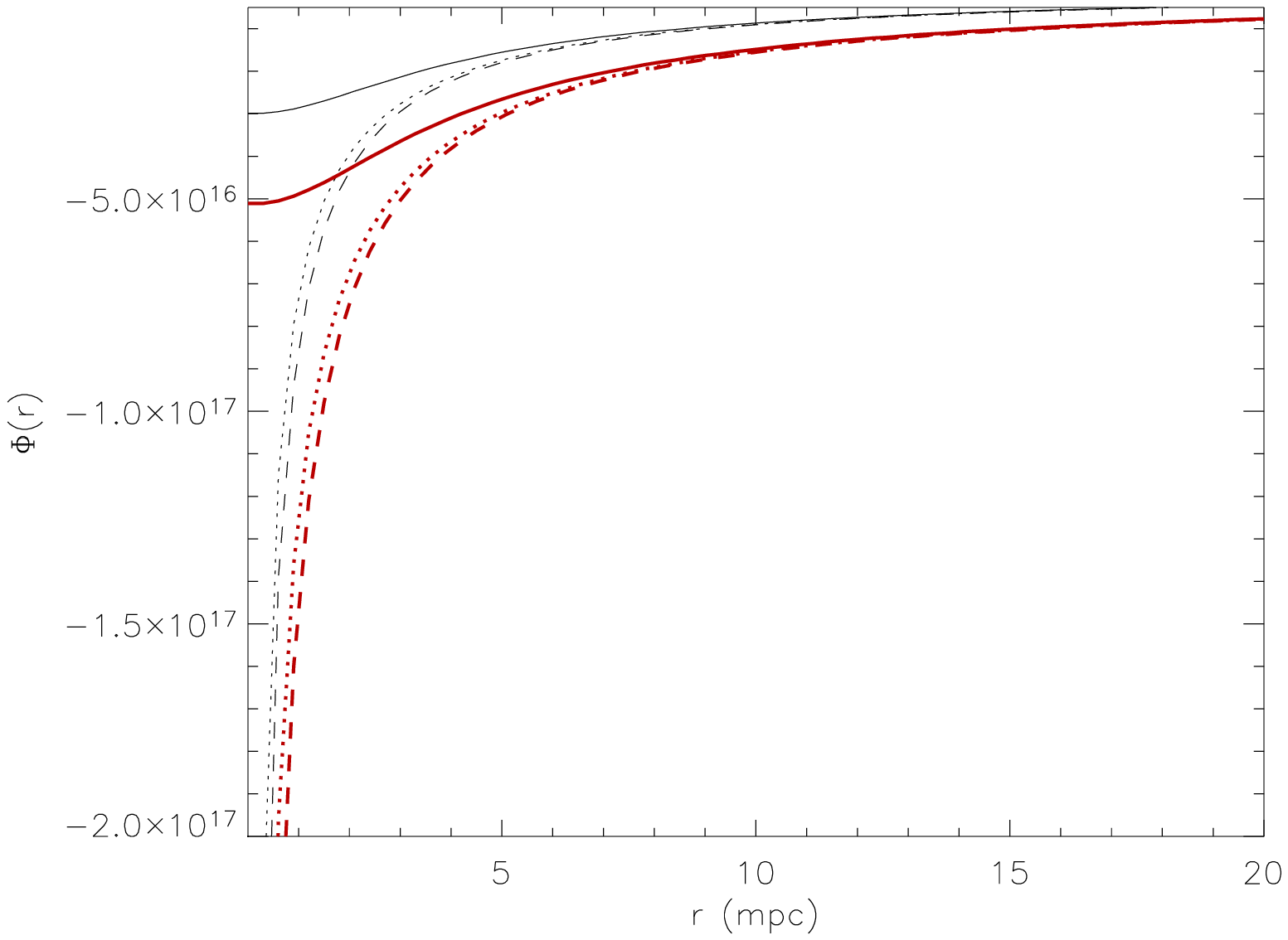}\qquad
\end{center}
\caption{The gravitational potential at distance $r$ as due to the
mass $M(r)$ is shown for different fractions $\lambda_{BH}$ of the
total mass $M$. Solid, dotted and dashed lines correspond to
$\lambda_{BH}= 0$, $\lambda_{BH}= 0.7$ and $\lambda_{BH}=0.9$,
respectively. Thick red lines have been obtained for $r_c=3$ mpc
while thin black lines are for $r_c=5.8$ mpc.} \label{potential}
\end{figure}
In Figure \ref{potential}, the gravitational potential $\Phi_N(r)$
due to the mass density distribution in eq. (\ref{totaldensity})
is given for selected values of $\lambda_{BH}$.

According to GR, the motion of a test particle can be fully
described by solving the geodesic equations. Under the assumption
that the matter distribution is static and pressureless, the
equation of motion of the test particle becomes (see e.g. Weinberg
\citeyear{Weinberg72})
\begin{equation}
\frac{d\textbf{v}}{dt}\simeq-\nabla(\Phi_N +2\Phi_N
^2)+4\textbf{v}(\textbf{v}\cdot \nabla)\Phi_N -v^2\nabla \Phi_N~.
\end{equation}
For a spherically symmetric mass distribution \footnote{We would
like to mention that the dynamical state of the region around Sgr
A$^*$ is known to be complex, with a significant population of
young stars of unclear origin making the assumption of an
undisturbed spherical cluster likely uncorrect. Considering the
effects caused by a non spherically symmetric mass distribution
makes the passage to an equation similar to eq. (\ref{setode}) not
analytically solvable. This problem will be addressed in a
subsequent work.} with a density profile given by eq.
(\ref{plummer1}) and for a gravitational potential given by eq.
(\ref{gravitationalpotential}), the previous relation becomes (see
for details Rubilar et al. \citeyear{Rubilar01})
\begin{equation}
\frac{d\textbf{v}}{dt}\simeq-\frac{GM(r)}{r^3}\left[\left(1+\frac{4\Phi_N}{c^2}+
\frac{v^2}{c^2}\right)\textbf{r}-\frac{4\textbf{v}(\textbf{v}\cdot\textbf{r})}{c^2}\right]~,
\label{setode}
\end{equation}
$\textbf{r}$ and $\textbf{v}$ being the radius vector of the test
particle (with respect to the center of the stellar cluster) and
the velocity vector, respectively. Once the initial conditions for
distance and velocity are given, the orbit of a test particle can
be found by solving the set of ordinary differential equations in
eq. (\ref{setode}) numerically.

Now consider the S2 star, which is moving around the central
distribution of matter on an elliptic orbit of semi-major axis $d$
and eccentricity $e$ in the Newtonian approximation. We take a
frame with the origin in the GC, $X$-$Y$ plane on the orbital
plane and the $X$ axis pointing toward the periastron of the
orbit. Hence, we can choose the Newtonian initial conditions to be
(see e.g. Smart (\citeyear{smart}))
\begin{eqnarray}
r_x^0&=&d(1+e)~, \nonumber \\
r_y^0&=&0~,\label{cond1}
\end{eqnarray}
and
\begin{eqnarray}
v_x^0&=&0,\nonumber \\
v_y^0&=&\sqrt{GM(r_x^0)\left[\frac{2}{d(1+e)}-\frac{1}{d}\right]}~.
\label{cond2}
\end{eqnarray}

\begin{figure*}[htbp]
\vspace{0.2cm}
\begin{center}
$\begin{array}{c@{\hspace{0.1in}}c@{\hspace{0.1in}}c}
\epsfxsize=2.0in \epsfysize=2.0in \epsffile{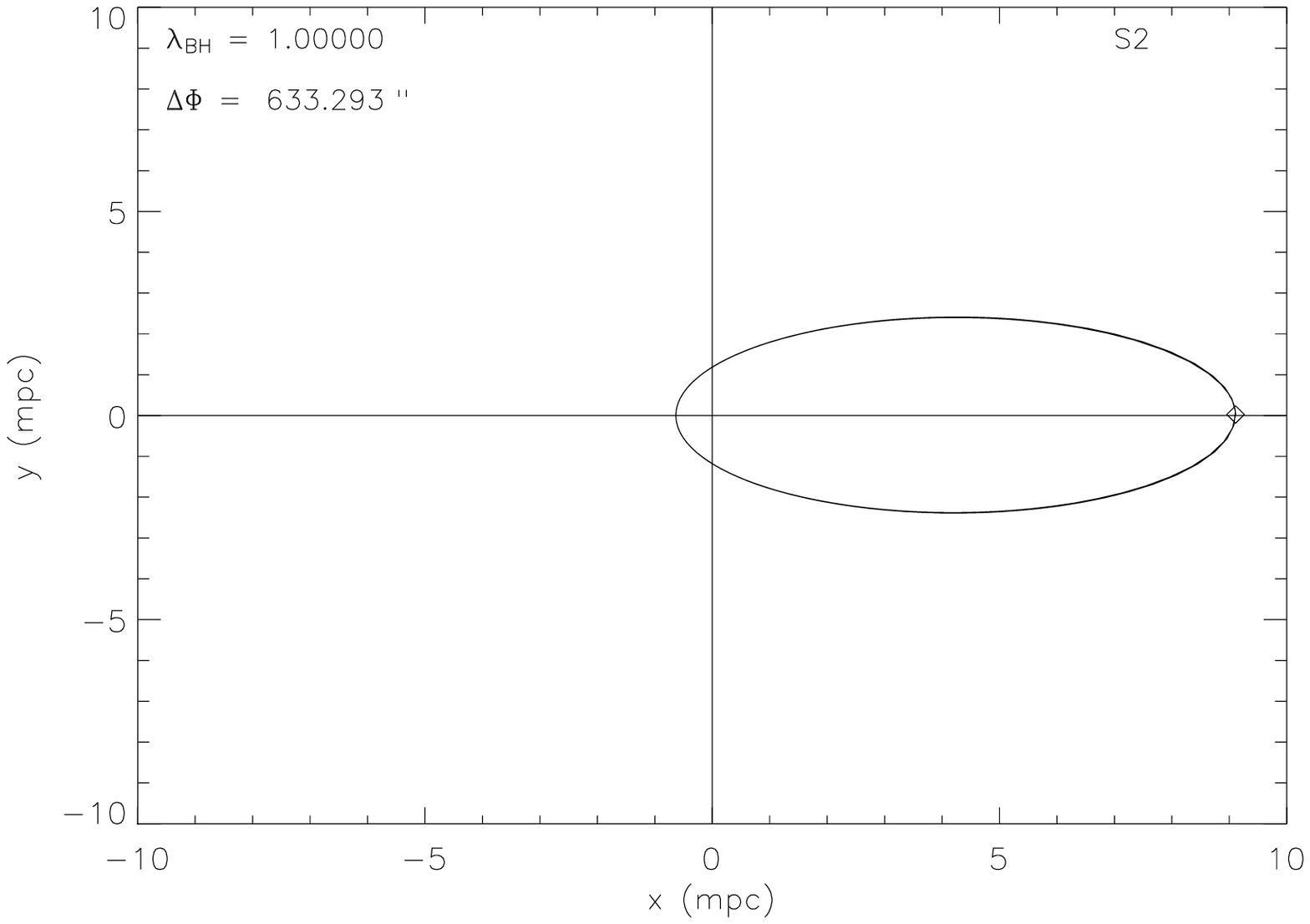} &
\epsfxsize=2.0in \epsfysize=2.0in \epsffile{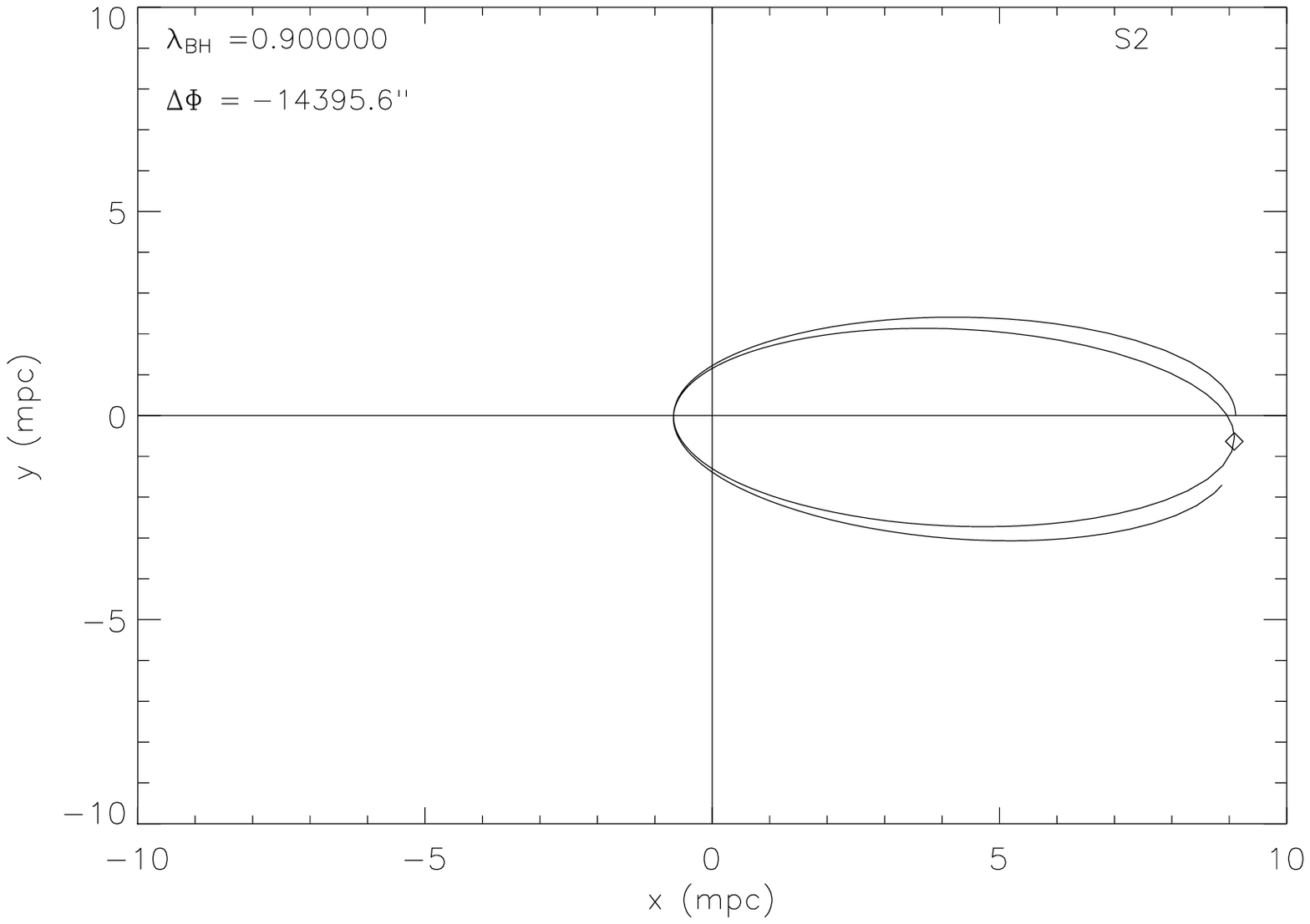} &
\epsfxsize=2.0in \epsfysize=2.0in \epsffile{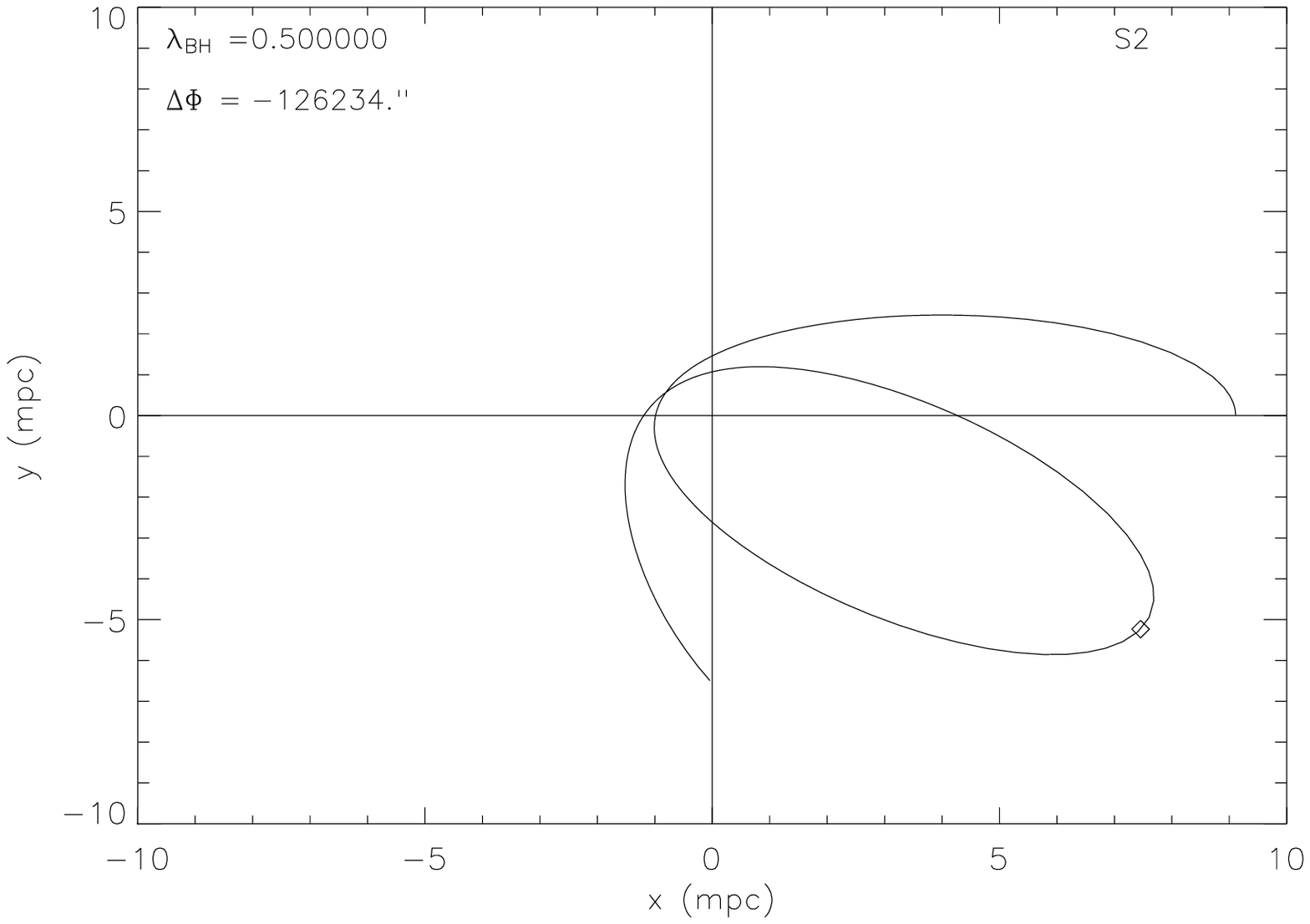} \\
\end{array}$
\end{center}
\caption{Post Newtonian orbits for different values of the black
hole mass  fraction $\lambda_{BH}$ are shown for the S2 star
(upper panels). Here, we have assumed that the Galactic central
black hole is surrounded by a stellar cluster whose density
profile follows a Plummer model with $\alpha=5$ and a core radius
$r_c\simeq 5.8$ mpc. The periastron shift values in each panel is
given in arcseconds.} \label{orbite}
\end{figure*}

\begin{figure*}[htbp]
\vspace{0.2cm}
\begin{center}
$\begin{array}{c@{\hspace{0.1in}}c@{\hspace{0.1in}}c}
\epsfxsize=2.0in \epsfysize=2.0in \epsffile{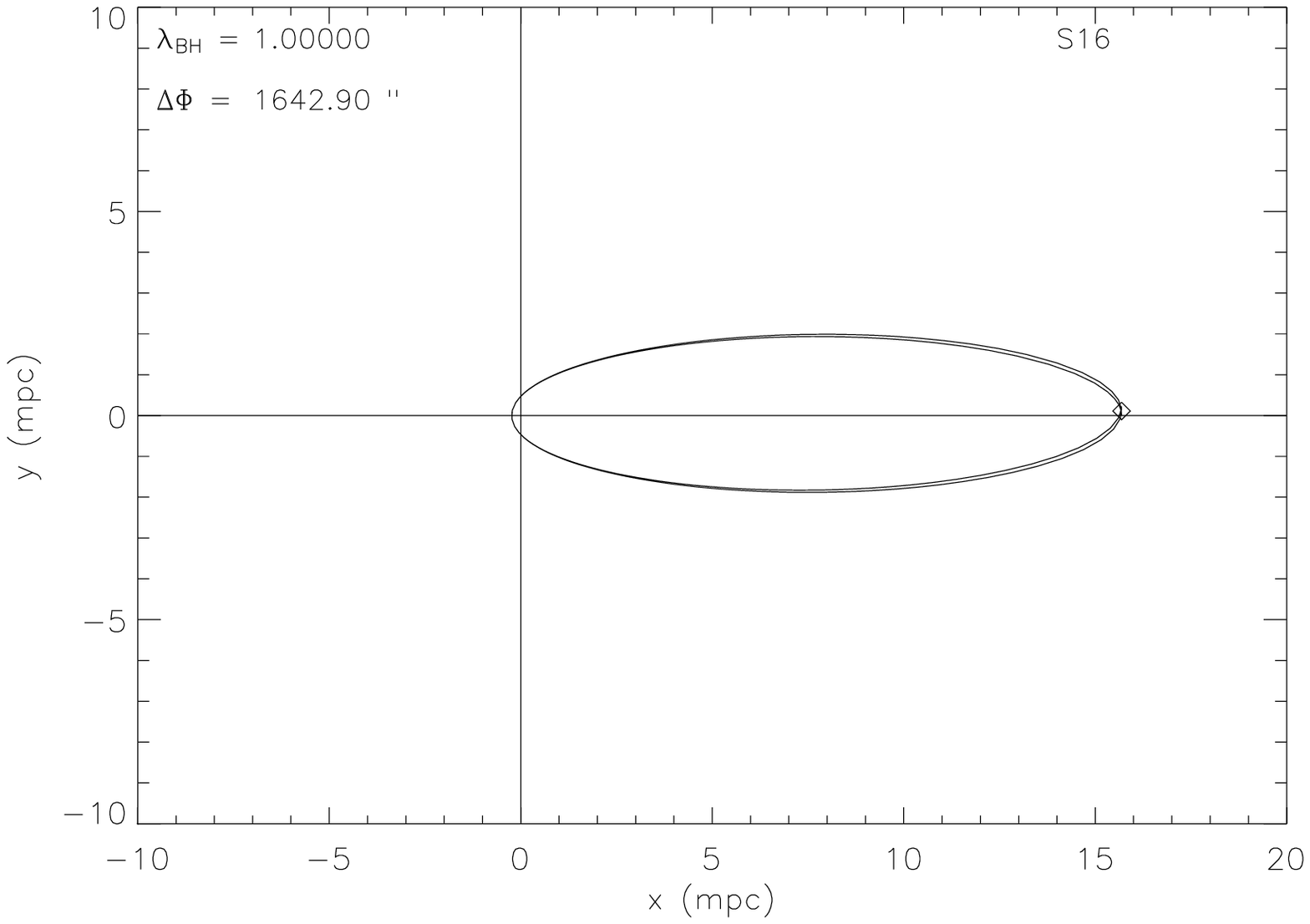} &
\epsfxsize=2.0in \epsfysize=2.0in \epsffile{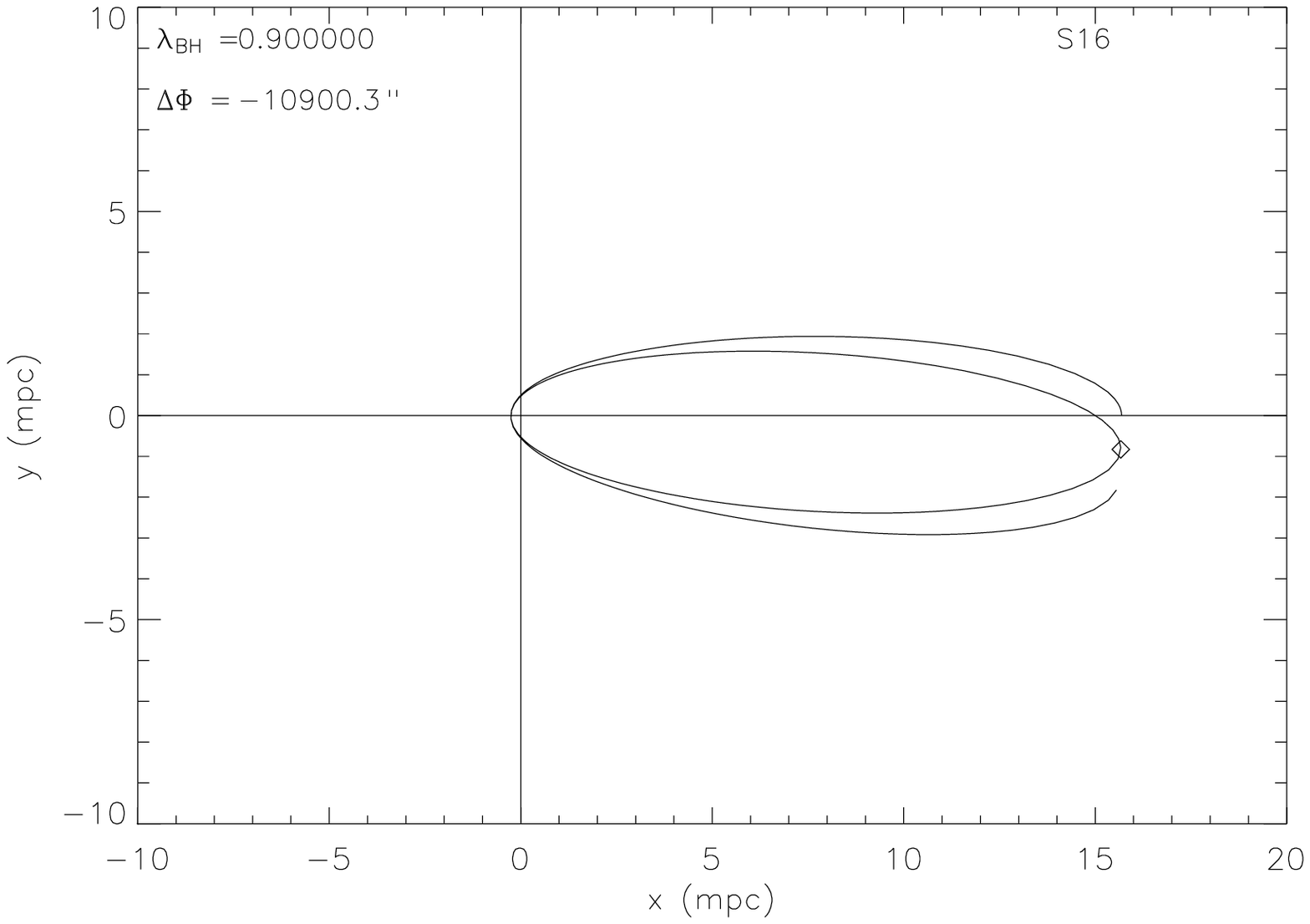} &
\epsfxsize=2.0in \epsfysize=2.0in \epsffile{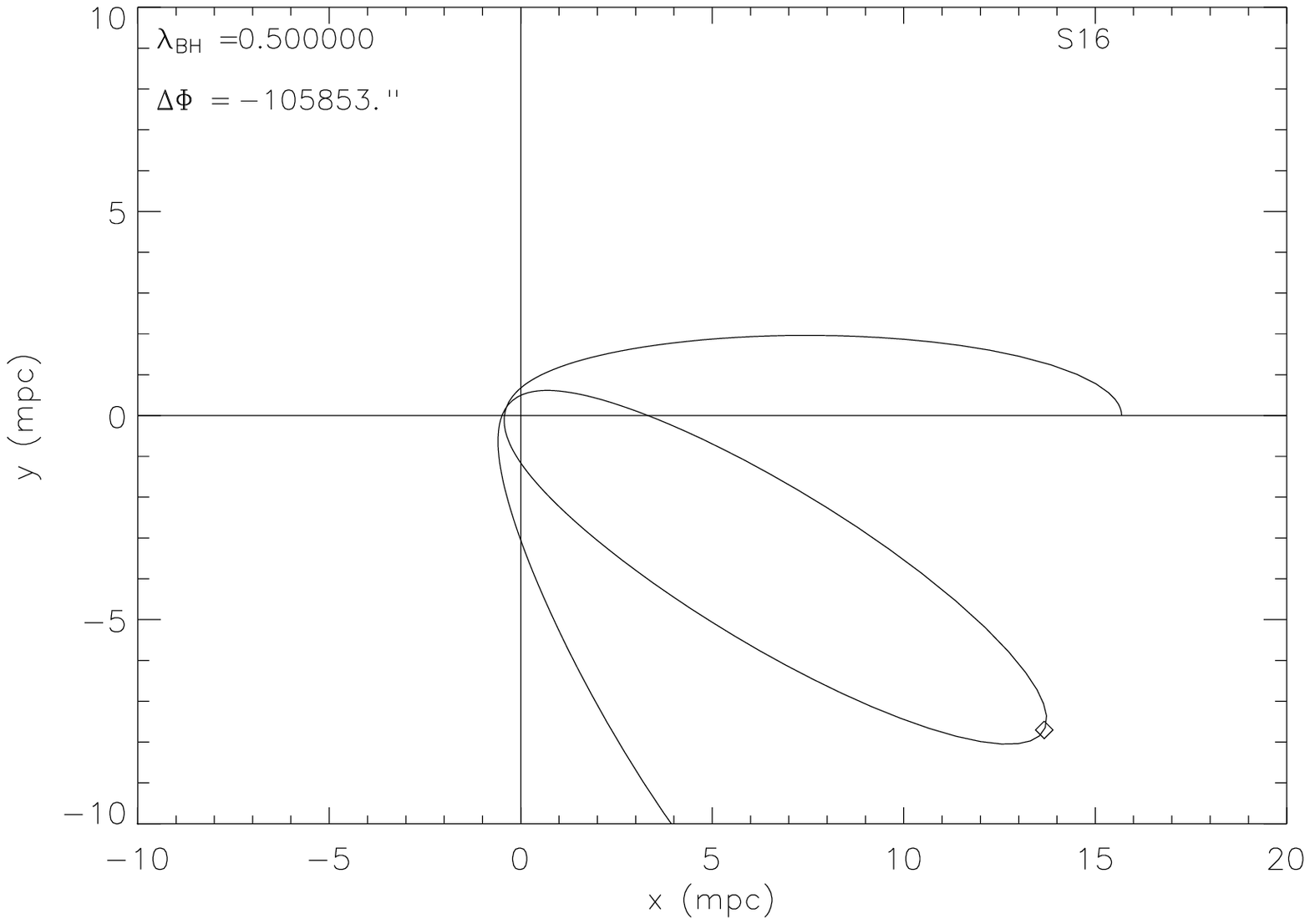} \\
\end{array}$
\end{center}
\caption{The same as in Figure \ref{orbite} but for the S16--Sgr
A$^*$ binary system. In this case, the binary system orbital
parameters were taken from Ghez et al. (\citeyear{Ghez05})
assuming for the S16 mass a conservative value of $\simeq 10$
M$_{\odot}$.} \label{orbite2}
\end{figure*}

For the S2 star, $d$ and $e$ given in the literature are 919 AU
and 0.87 respectively. They yield the orbits of the S2 star for
different values of the black hole mass fraction $\lambda _{BH}$
shown in Figure \ref{orbite}. The Plummer model parameters are
$\alpha =5$, core radius $r_c\simeq 5.8$ mpc. Note that in the
case of $\lambda _{BH}=1$, the expected (prograde) periastron
shift is that given by eq. (\ref{schshift}), while the presence of
the stellar cluster leads to a retrograde periastron shift. For
comparison, the expected periastron shift for the S16 star is
given in Figure \ref{orbite2}. In the latter case, the binary
system orbital parameters were taken from Sch\"odel et al.
(\citeyear{Schoedel03}) assuming also for the S16 mass a
conservative value of $\simeq 10$ M$_{\odot}$.

\begin{figure*}[htbp]
\vspace{0.2cm}
\begin{center}
$\begin{array}{c@{\hspace{0.1in}}c@{\hspace{0.2in}}c}
\epsfxsize=2.8in \epsfysize=2.8in \epsffile{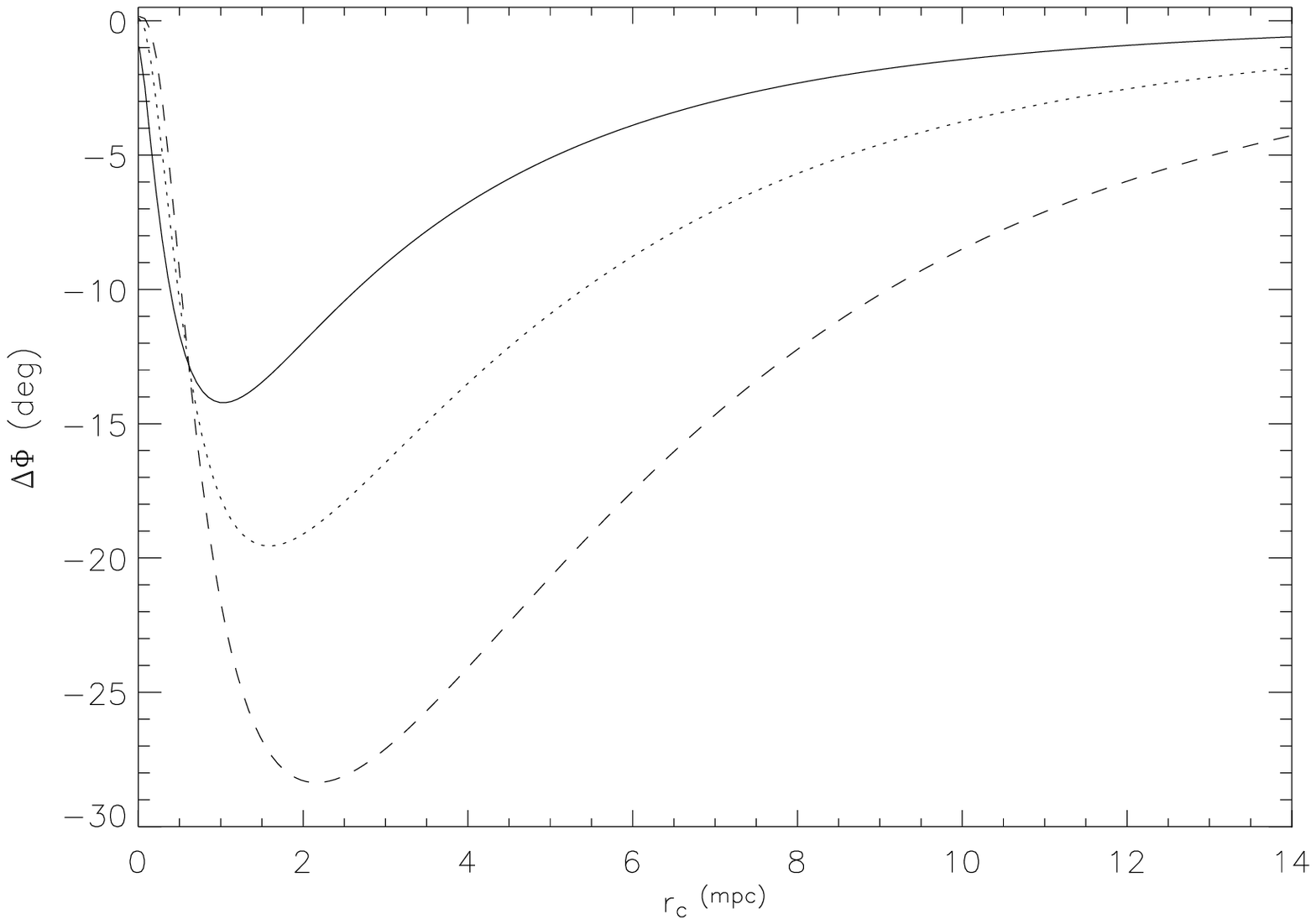} &
\epsfxsize=2.8in \epsfysize=2.8in \epsffile{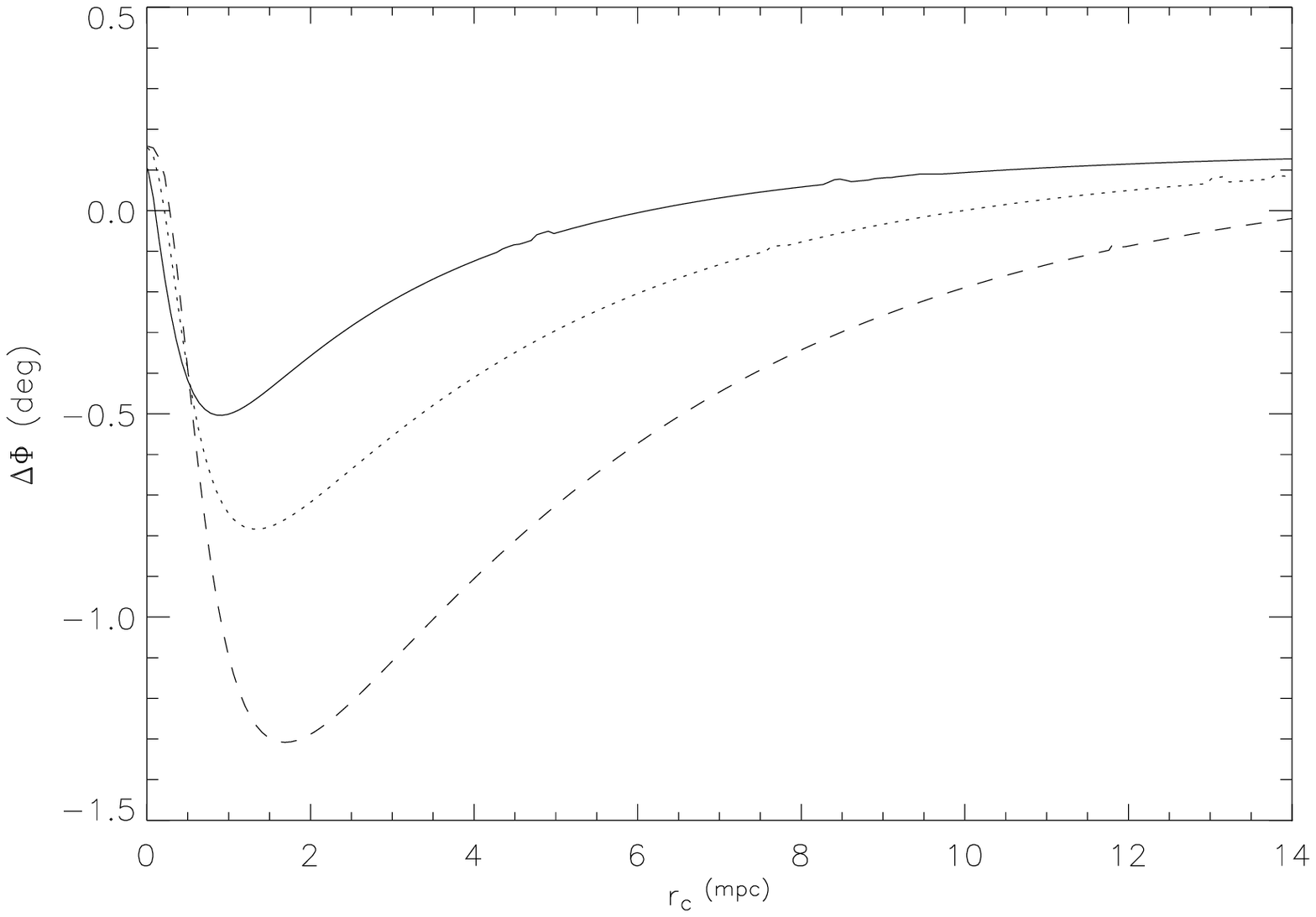} \\
\end{array}$
\end{center}
\caption{The expected S2 periastron shift as a function of the
stellar cluster core radius is shown. Here we have assumed a
Plummer density profile for the stellar cluster. Dashed, solid and
dotted lines correspond to $\alpha=4,~5$ and $6$, respectively.
The black hole mass  fraction has been fixed to $\lambda_{BH}=0.8$
(left panel) and $\lambda_{BH}=0.99$ (right panel), respectively.
Note the existence of a maximum approximately corresponding to the
S2 semi-major axis.} \label{shiftvscore}
\end{figure*}

In Figure \ref{shiftvscore} the S2 orbital shift $\Delta \Phi$ is
given as a function of the stellar cluster core radius $r_c$, for
different power law index values ($\alpha=4$ dashed line,
$\alpha=5$ dotted line and $\alpha=6$ solid line). In the left
panel, the black hole mass fraction is $\lambda_{BH}=0.8$ in order
to compare with Rubilar et al. (\citeyear{Rubilar01}) results,
while the right panel shows the $\lambda_{BH}=0.99$ case. Note
that for extremely compact clusters, $\Delta \Phi$ is quite small.
The same is true for large enough core radii, corresponding to
matter density profiles almost constant within the S2 orbit.

Figures \ref{orbite} and \ref{shiftvscore} show that the expected
S2 periastron shift depends strongly on the total mass of the
cluster. In particular, the shift due to the cluster is opposite
in sign (retrograde motion) to the relativistic effect due to the
black hole in the GC. Moreover, for each value of the cluster mass
and power law index, there exist two density profiles
(corresponding to two particular core radii) which have total
shift almost zero, implying that the periastron advance due to the
cluster is equal (in magnitude) to the periastron shift due to the
black hole. A numerical analysis shows that the transition from a
prograde shift (due to the black hole) to retrograde shift (due to
the extended mass) occurs at $\lambda_{BH} \simeq 0.9976$,
$0.9986$ and $0.9990$ for $\alpha =4$, $5$ and $6$, respectively.
This means that a small fraction of mass in the cluster
drastically changes the overall shift.

We would like to note that the assumption of the Plummer model to
describe the mass density distribution of the stellar cluster
around the central black hole is an oversimplification. Indeed,
one expects that in presence of a central black hole, the stellar
profile should follow a Bachall-Wolf law with density distribution
$\rho_c(r) \propto r^{-7/4}$ \citep{bw,binneytremaine} at least up
to $\tilde{r}_H\ll r_H$, where $r_H=GM_{BH}/\sigma_*^2\simeq 0.5$
pc is the radius of the black hole influence sphere. In the
following, we call $\tilde{r}_H$ the distance ($\ll r_H$) up to
which the cluster mass density follows the Bachall-Wolf law.

In order to study the effect of such a cusp on the expected S2
periastron shift, we consider three different cases {\it a)} the
cusp is entirely contained within the S2 periastron distance
$R_{S2}$ (i.e. $\tilde{r}_H\le R_{S2}$), {\it b)} the cusp extends
beyond the S2 periastron distance (thus making the S2 star move in
a mass gradient) and {\it c)} the stellar density profile follows
a cusp law up to the distance $\tilde{r}_H$ from the center and a
Plummer law for $r\geq \tilde{r}_H$. In cases {\it a)} and {\it
b)} all stars are in a cusp density profile. In any case we
require that the total mass enclosed within $4.87\times 10^{-3}$
pc is $M\simeq 3.67\times 10^6 ~M_{\odot}$.

In case {\it a)}, the total S2 periastron shift is just the sum of
the shift due to the black hole and the shift caused by the
stellar cusp (that contributes with the same sign). Hence, the S2
shift turns out to be $\Delta \Phi \simeq 0.17$ degree per
revolution.

\begin{figure}[h]
\begin{center}
\includegraphics[scale=0.50]{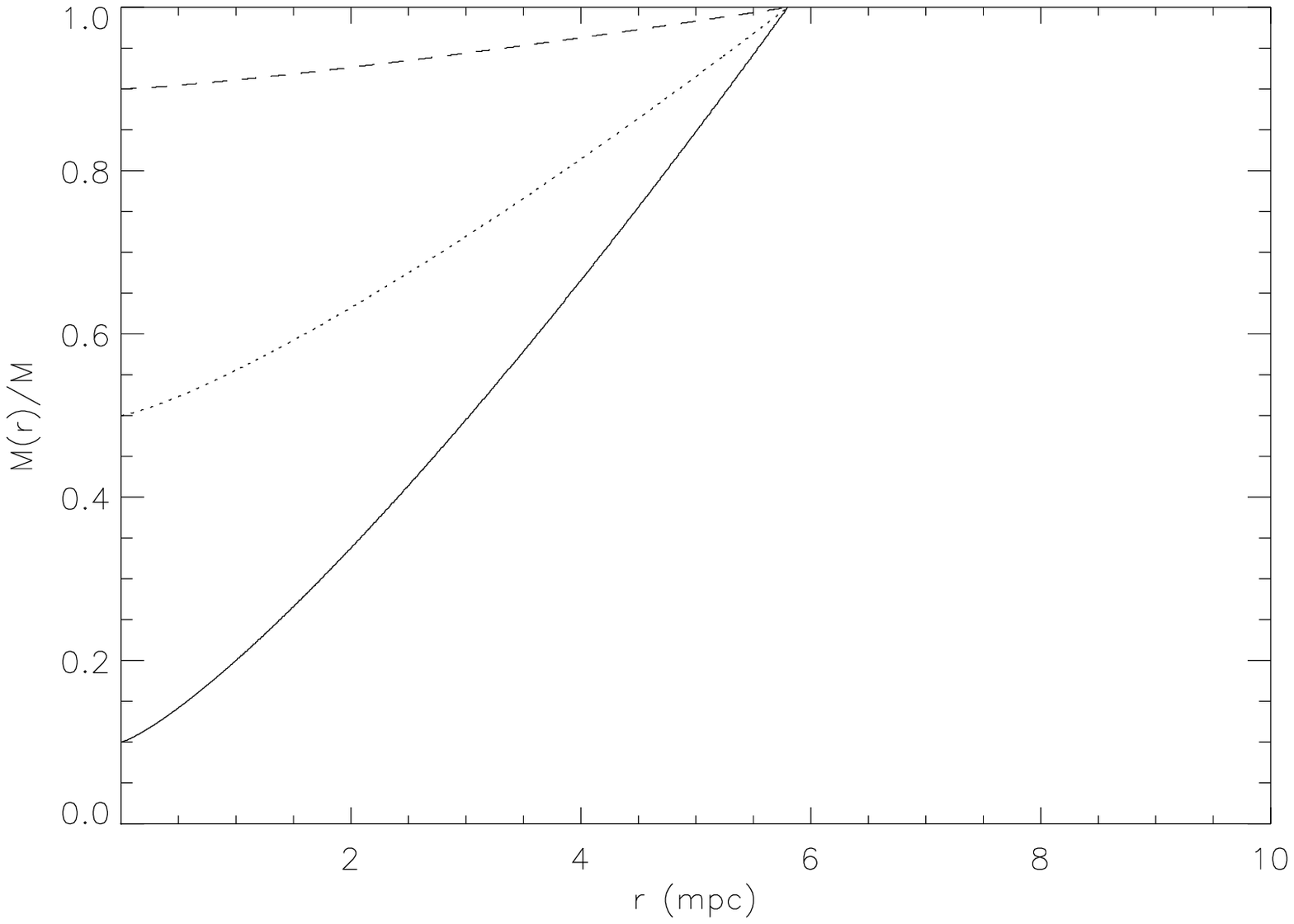}\qquad
\end{center}
\caption{The mass enclosed within the distance $r$ is shown for
different fractions $\lambda_{BH}$ of the total mass $M$ contained
within the S2 orbit. Solid, dotted and dashed lines correspond to
$\lambda_{BH}= 0.1$, $\lambda_{BH}= 0.5$ and $\lambda_{BH}=0.9$,
respectively. The stellar cluster is assumed to follow an
$r^{-7/4}$ density profile.} \label{cusp1}
\end{figure}

\begin{figure}[h]
\begin{center}
\includegraphics[scale=0.50]{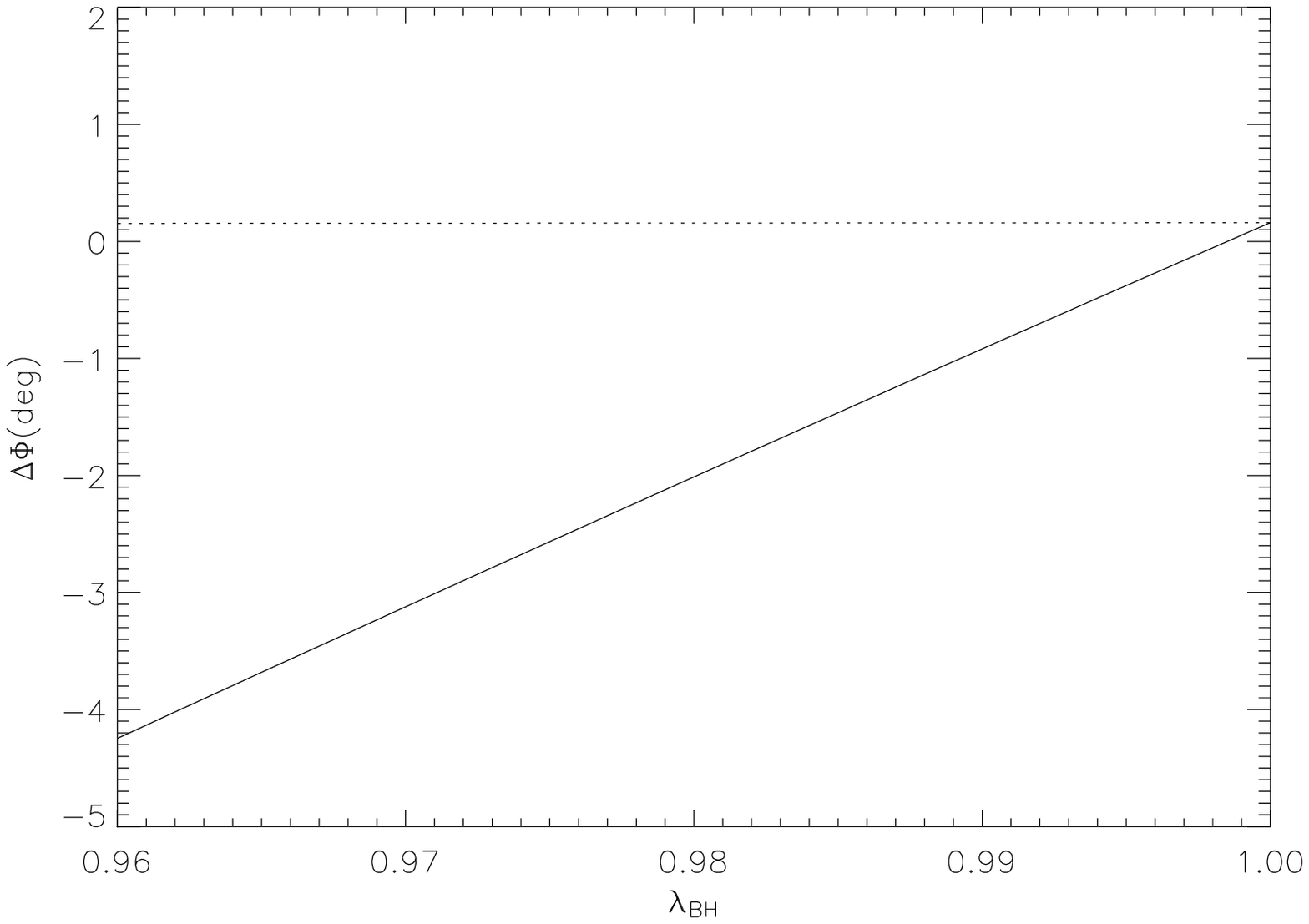}\qquad
\end{center}
\caption{The expected S2 periastron shift as a function of the
mass ratio parameter $\lambda_{BH}$. Solid and dashed lines
correspond to the S2 shift due to the black hole and to the
stellar cusp, respectively. We note that the shift due to the
stellar cusp is independent on the $\tilde{r}_H$ value, that, in
this case, has been assumed to be larger than the S2 semi-major
axis (case {\it b}).} \label{cusp2}
\end{figure}

In case {\it b)}, by requiring that the total mass enclosed within
$4.87\times 10^{-3}$ pc is $M\simeq 3.67\times 10^6 ~M_{\odot}$,
we find that the dependence of the cusp mass and the induced S2
periastron shift on $\tilde{r}_H$ vanishes. Indeed, in Figure
\ref{cusp1}, we give the mass enclosed within the distance $r$ for
different values of $\lambda_{BH}$. Solid, dotted and dashed lines
correspond to $\lambda_{BH}= 0.1$, $\lambda_{BH}= 0.5$ and
$\lambda_{BH}=0.9$, respectively. Figure \ref{cusp2} shows the
expected S2 periastron shift as a function of $\lambda_{BH}$. As
noted before, the shift due to the cluster is opposite in sign
with respect to that due to the black hole. Moreover, for
$\lambda_{BH}\simeq 0.998$ the total shift turns out to be zero
since the contributions of the black hole and the cluster cancel
out. It is noticing that, since in the case of cusp profiles the
density gradient is larger than in the case of a usual
($\alpha=5$) Plummer model, the value of the S2 periastron shift
gets generally larger values. Only if the Plummer core radius is
around $2$ mpc the resulting S2 periastron shifts are comparable
in both cases.
We have then considered the superposition of a Plummer model and a
Bahcall-Wolf profile (case {\it c}) extended up to $\tilde{r}_H$
such as the cusp density at  $\tilde{r}_H$ equals that of the
Plummer model at the same distance. Here, if  $\tilde{r}_H\ll
R_{S2}$ the S2 periastron shift will be practically equal to that
caused by the Plummer model (see right panel in Fig. 6) since in
this case the cusp will have a minor influence. On the contrary,
for an extended cusp ($\tilde{r}_H\gg R_{S2}$), the cusp effect on
the S2 periastron shift will dominate reconciling with case $b$.

As a last point, we mention that we have also considered the
effect due to an extrapolation of the observed stellar density
profile - the innermost point of which is the S2 star at a
distance of 0.1\arcsec - within $R_{S2}$. Following
\citet{Genzel03b} and assuming  a cusp stellar density profile, we
find that the enclosed mass is in the range 30-300 $M_{\odot}$
(for a constant mass density or a power law with index
$\gamma=1.4$). Therefore, the cusp effect on the S2 periastron
shift is negligible since the corresponding $\lambda_{BH}$ is
always greater than 0.99992. However, we caution that the case
under investigation in the present paper is different with respect
to \citet{Genzel03b} since we are assuming that a fraction of the
mass contained within $R_{S2}$ may be in a stellar cluster. Hence,
the cluster mass content may be larger, thus providing a stronger
effect on the S2 periastron shift.

\section{Tightening mass limits of Sgr~A$^*$}

We know that the mass of Sgr~A$^*$ within the S2 orbit is
$3.67\times 10^6$ M$_{\odot}$ to a high accuracy. Though there is
nothing definite known about the mass distribution, there is
strong reason to believe that there is a black hole of several
solar masses, possibly surrounded by a significant cluster. In
principle the cluster mass could dominate over the black hole, be
comparable to it or be dominated by it. That there {\it is} a
cluster is highly likely on account of the large number of stars
observed near Sgr~A$^*$. Though these lie outside the S2 orbit,
many stars so far unseen probably do lie within the orbit as well.
In this section we use current data on the Brownian motion of
Sgr~A$^*$ and the evaporation time for the putative cluster to put
limits on the cluster mass and hence on the black hole mass.

Chatterjee, Hernquist and Loeb (\citeyear{chatterje}) have
developed a simple model to describe the dynamics of a massive
black hole surrounded by a dense stellar cluster. The total force
acting on the black hole is separated into two independent parts,
one of which is the slowly varying force due to the stellar
ensemble and the other the rapid stochastic force due to close
stellar encounters. In the case of a stellar system with a Plummer
distribution, the motion of the black hole is similar to that of a
Brownian particle in a harmonic potential. Thus the black hole
one-dimensional mean-square velocity is given by
\begin{equation}
<v_x^2>=\frac{2}{9}\frac{GM_{CL}m_*}{r_cM_{BH}}~, \label{eqvel1}
\end{equation}
where it has been assumed that the cluster is composed of objects
with equal mass, $m_{*}$. For a Plummer ($\alpha$ = 5) stellar
cluster, the total mass within $R$ is
\begin{equation}
M(R)=M_{BH}+\frac{M_{CL}R^3}{(R^2+r_c^2)^{3/2}}~. \label{eqmass1}
\end{equation}

\begin{figure}[h]
\begin{center}
\includegraphics[scale=0.55]{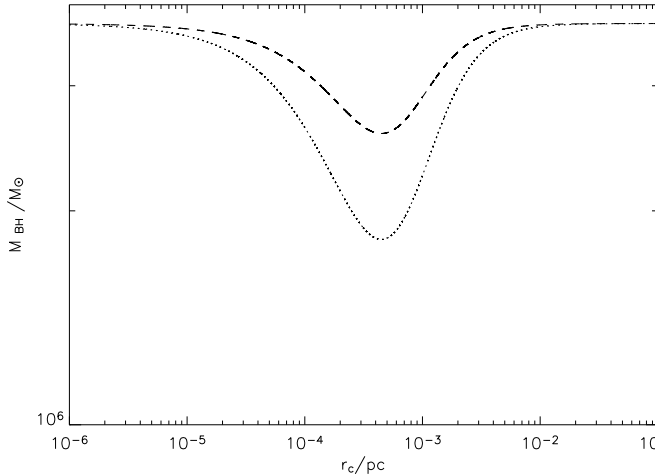}\qquad
\end{center}
\caption{The minimum black hole mass allowed by the Brownian
motion of Sgr~A$^*$ is given as a function of the stellar cluster
core radius for the different black hole proper motion velocities.
We assume that a total mass $M\simeq 3.67\times 10^6$ M$_{\odot}$
is contained within $R_{S2}=4.87$ mpc. Dashed and dotted lines
have been obtained for velocities of $1.3$ km ${\rm s}^{-1}$ and
$2$ km $\rm{s}^{-1}$, respectively. For each given curve only the
region above it is allowed.}\label{figlimiti1}
\end{figure}

Since $<v_x^2>$ is less than a certain maximum value
$<v_x^2>_{max}$, from eqs. (\ref{eqvel1}) and (\ref{eqmass1}) one
obtains
\begin{equation}
M_{BH} > M(R) \left\{1+\frac{9}{2} \left[ <v_x^2>_{max} \frac{r_c
R^3}{G (R^2+r_c^2)^{3/2}m_* } \right]\right\}^{-1}~,\label{eqlim1}
\end{equation}
the right hand side corresponding to a minimum black hole mass, as
constrained by the Brownian motion of the central black hole. In
Figure \ref{figlimiti1} the minimum black hole mass allowed by the
Brownian motion of Sgr~A$^*$ is given as a function of the stellar
cluster core radius, for two different proper motion velocities of
the black hole: 1.3 km s$^{-1}$ (dashed lines), and 2 km s$^{-1}$
(dotted lines). The total mass contained within $R=0.01$ pc of
Sgr~A$^*$ has been taken to be $M\simeq 3.67\times 10^6$
M$_{\odot}$.

Chatterjee, Hernquist and Loeb (\citeyear{chatterje}) derived an
evaporation time for a cluster, but concentrated on the large
scale cluster $r_c\simeq$ 10 pc about Sgr~A$^*$, and hence assumed
that M$_{CL}$ $\gg$ M$_{BH}$. On the other hand Rauch and Tremaine
(\citeyear{rauch}) and Mouawad et al. (\citeyear{moawad}) consider
only the region interior to the orbit of S2 and assume M$_{CL}$
$\ll$ M$_{BH}$.

We need to allow for all possibilities while considering the
cluster interior to the orbit of S2, including M$_{CL}\simeq$
M$_{BH}$. For this purpose we consider a cluster of core radius
$r_c$ and mass M$_{CL}$ = M $-$ M$_{BH}$. We now need to obtain
the generalization of the formula of Chatterjee, Hernquist and
Loeb (\citeyear{chatterje}) for the median relaxation time in this
more general situation. For this purpose, as usual, we assume that
the cluster consists of components of the same mass $m_*$ and
evaluate the crossing time in the usual way to obtain the general
median relaxation time
\begin{equation}
T_{r}=\frac{0.14(1.3~r_cM)^{3/2}}{\sqrt{G}M_{CL}m_*log(0.4M/m_*)}.
\end{equation}
It is easy to verify that in the approximation M$_{CL}$ $\gg$
M$_{BH}$ we recover the formula of Chatterjee, Hernquist and Loeb
(\citeyear{chatterje}) and in the approximation M$_{CL}$ $\ll$
M$_{BH}$ we recover the formula of Rauch and Tremaine
(\citeyear{rauch}).The evaporation time is then $T_{evap}\simeq
300~T_{r}$ (\citealt{binneytremaine}, p.525).

One can assume different ``reasonable" values of the time that the
cluster would have been in existence and hence use the evaporation
time to further limit the black hole mass in the GC. It is clear
that 10$^8$ years = 0.1 Gyr is less than the minimum value that
could be regarded as reasonable, 1 Gyr is more reasonable and 10
Gyr is likely to be a good value to take. The results are given in
Table \ref{tab3} for $m_*$ = 1 M$_{\odot}$. Note that the tightest
bound gives a very stringent upper limit of
$9\times10^4$M$_{\odot}$ on the cluster mass. Also note that the
value decreases if the average $m_*$ is taken to be larger.
\begin{table*}
\begin{center}
\begin{tabular}{|c|c|c|c|c|}
\hline \rule{0pt}{3ex} $\rm{ T_{evap} (Gyr)}$ & ${\rm r_c^{1.3}
(mpc)}$ & ${\rm \lambda_{BH}^{1.3}}$ & ${\rm r_c^{2.0}
(mpc)}$ & ${\rm \lambda_{BH}^{2.0}}$ \\
\hline
\rule{0pt}{3ex}$0.1$  & $0.87$  & $0.762$ & $1.28$  & $0.645$ \\
\hline
\rule{0pt}{3ex}$1$  & $2.11$  & $0.919$ & $2.61$  & $0.876$ \\
\hline
\rule{0pt}{3ex}$10$  & $4.12$  & $0.975$ & $5.27$  & $0.964$ \\
\hline
\end{tabular}
\end{center}
\caption{The cluster core radius $r_c$ and minimum black hole mass
fraction $\lambda_{BH}$ for the limits obtained by
$<v>_{max}^2=1.3$ and 2.0, for $T_{evap}=0.1$, 1 and 10 Gyr.}
\label{tab3}
\end{table*}

\section{The spin of the black hole}

The periastron shift is the net contribution of the relativistic
retrograde shift due to the black hole and the Newtonian prograde
shift due to the surrounding cluster. Obviously, if the periastron
advance due to the stellar cluster were known, the contribution of
periastron advance due to the black hole could be obtained by
subtracting from the measured quantity. The question arises
whether the information obtained would be adequate to obtain both
the black hole mass and spin parameters. Though we can put
reasonably sharp bounds on the stellar cluster about the black
hole, is it good good enough for our purpose? If so, we could use
eq.(\ref{kershift}) to obtain the spin of the black hole for
different values in the possible range for the periastron shift.
It is easy to see from Fig.\ref{shiftvscore} that for
$\lambda_{BH} = 0.99$ and allowing for the maximum range of
unknown values of $\alpha$ and $r_{c}$ the $1.8\times10^3<-\Delta
\phi<4.7\times10^3$ or $1.9\times10^{-3}<-\Delta
\phi_E<4.7\times10^{-3}$. For the sharpest limit obtained,
$\alpha=5$, and Brownian motion 1.3 km s$^{-1}$, $\lambda_{BH} =
0.975$, we find that $\Delta \phi \simeq -4.47\times10^3$ or
$\Delta \phi_E \simeq -4.7\times10^{-3}$. For Brownian motion 2.0
km s$^{-1}$, $\lambda_{BH} = 0.964$, $\Delta \phi \simeq
-5.75\times10^3$ or $\Delta \phi_E \simeq -6.0\times10^{-3}$. This
is a factor of 5 {\it less} than the effect of the spin. Hence
this method {\it cannot} be used to determine the spin. For this
we need the cluster parameter values, rather than upper limits for
them. Alternatively, one would need to rely on the retrolensing
method suggested earlier (\citealt{depaolis4,depaolis5}).

\section{Determination of cluster parameters}

Using the stronger (1$\sigma$) limit of 1.3 km $\rm{s}^{-1}$ and
the weaker (2$\sigma$) limit of 2.0 km $\rm{s}^{-1}$ to limit the
Brownian motion of Sgr A$^*$ for our calculations and evaporation
times of 1 and 10 Gyr for the cluster, we obtained the minimum
black hole mass. For the stronger limits on the Brownian motion
and the evaporation time, it is $3.579\times 10^6$ M$_{\odot}$
corresponding to a $\lambda_{BH} \simeq 0.975$ for $\alpha =5$.
Our numerical analysis shows that the transition from a prograde
shift (due to the black hole) to a retrograde shift (due to the
extended mass assumed to be distributed with a Plummer density
profile) occurs at $\lambda_{BH} \simeq 0.9976$, $0.9986$ and
$0.9990$ for $\alpha =4$, $5$ and $6$, respectively. Hence, even a
small cluster around the central massive black hole limits the
possibility to observe and use the periastron shift of the S2
star.

\begin{figure}[h]
\begin{center}
\includegraphics[scale=0.55]{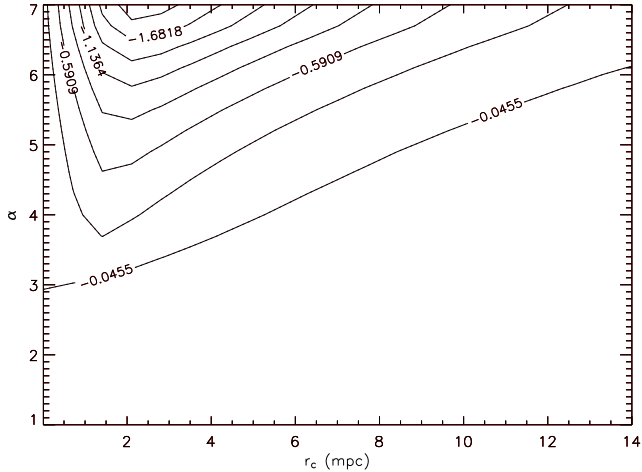}\qquad
\end{center}
\caption{The expected S2 periastron shift for $\lambda _{BH}=0.99$
for different values of both the core radius $r_c$ and power-law
index $\alpha$ for the considered Plummer density profile. Each
contour line corresponds to a given S2 periastron shift in degree
units. Note that a degeneracy occurs since there exist different
values of the power law index and core radius corresponding to the
same periastron shift.} \label{fig_par0}
\end{figure}

\begin{figure}[h]
\begin{center}
\includegraphics[scale=0.55]{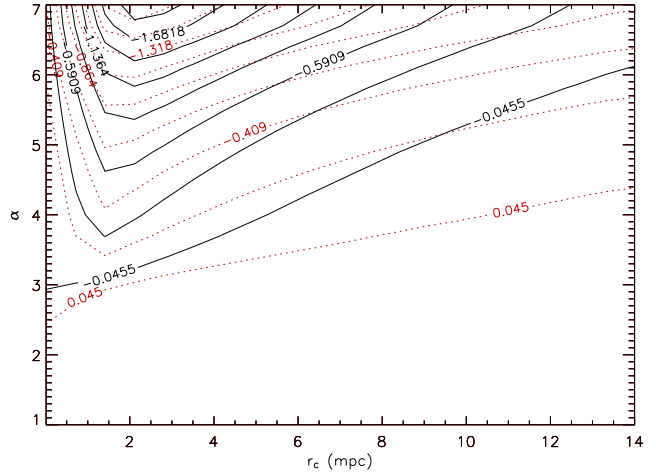}\qquad
\end{center}
\caption{The same as in Figure 8. Dotted lines show contours for
the S16 star. If the periastron shifts of both stars will be
measured in the future the intersection between the corresponding
contour lines will give information about the central stellar
density profile.} \label{fig_par1}
\end{figure}

\begin{figure}[h]
\begin{center}
\includegraphics[scale=0.55]{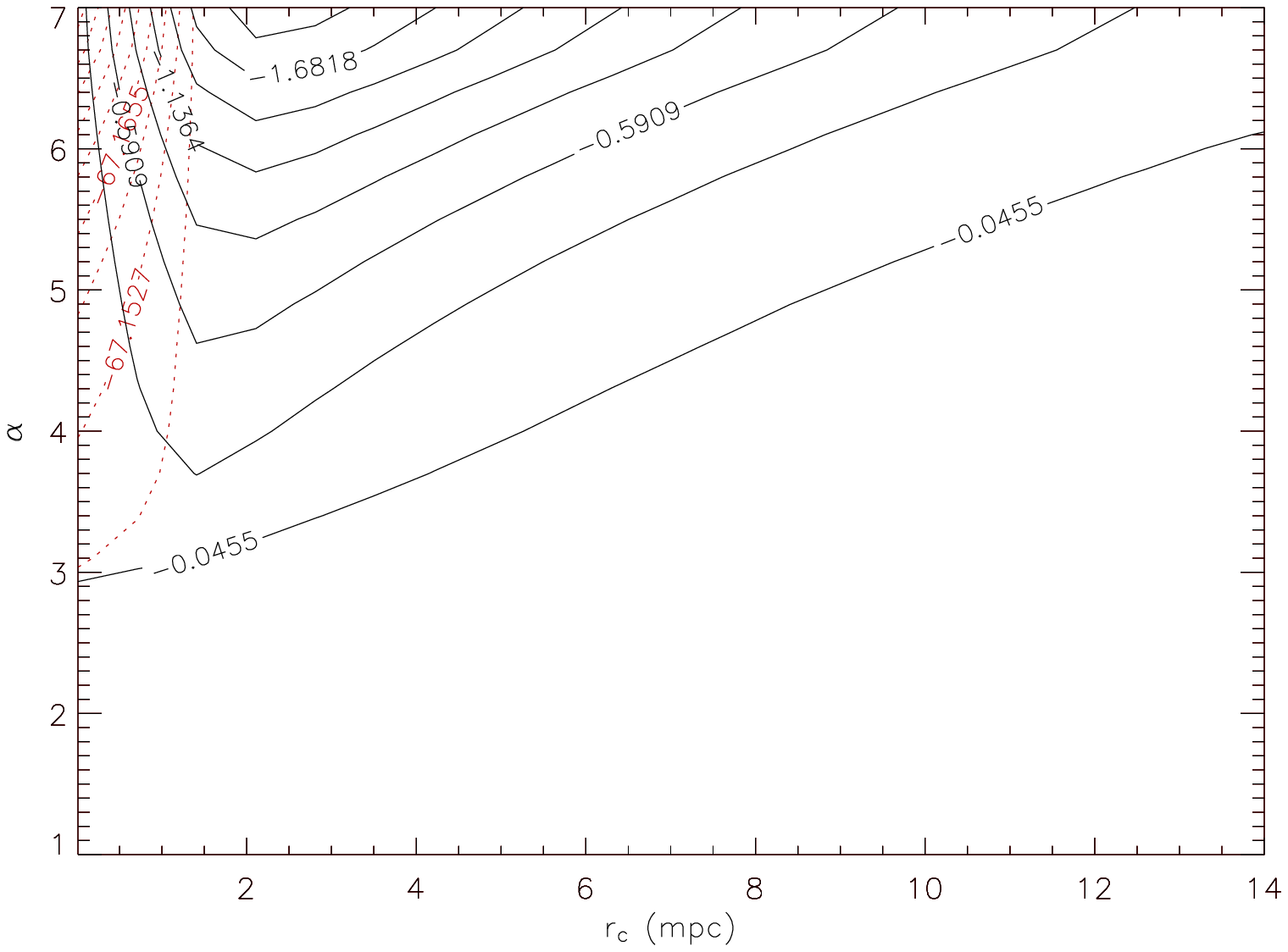}\qquad
\end{center}
\caption{The same as in Figure 8. Dotted lines show contours for a
star with orbit of $\simeq 0.87$ (the same as the S2 star) but
semi-major axis 3 times smaller with respect to the S2 one. If the
periastron shifts of both stars will be observed in the future the
intersection between the corresponding contour lines gives
information about the central stellar density profile.}
\label{fig_par2}
\end{figure}

Since we have modeled the star cluster density profile by a
Plummer model, the periastron shift contribution due to the
stellar cluster depends on three parameters: the central density
$\rho _0$ (or equivalently $\lambda _{BH}$); the core radius
$r_c$; and the power-law index $\alpha$. This degeneracy in the
determination of the stellar cluster parameters is due to the
measurement of the periastron shift of a single star. This is
easily seen by inspecting Figure \ref{fig_par0}, which has been
obtained for illustrative purposes for the S2 star by setting
$\lambda _{BH}=0.99$ and varying both the core radius $r_c$ and
power-law index $\alpha$ for the star cluster density profile.
Each contour line corresponds to a given S2 periastron shift in
units of degrees. To solve the parameter degeneracy and determine
the stellar cluster parameters (by studying the periastron advance
effect), the periastron shifts for at least three different stars
have to be measured with sufficient accuracy. Consider, for
example, the S16 star having an orbital period of $\simeq 36$ yr
and eccentricity $e\simeq 0.97$. Measuring its periastron shift
and comparing with the S2 result will give much tighter
information about the stellar cluster parameters. From Figure
\ref{fig_par1} it is evident that there are regions (intersections
between dashed and solid lines) in the $\alpha$-$r_c$ plane for
which one measures values of the periastron shift for the S2 and
S16 stars. Obviously, there could be (as yet unobserved) stars
with orbit apocenters comparable to S2, but with different
eccentricities (for example larger than 0.87) or stars closer to
the GC black hole than S2 or S16 stars. Monitoring their orbits
and measuring their periastron shifts will be extremely helpful in
reconstructing the cluster density profile. As an example, in
Figure \ref{fig_par2} we compare the expected S2 periastron shift
(solid lines obtained for $\lambda _{BH}=0.99$) with the
periastron shift of a star whose orbit has an eccentricity of
$\simeq 0.87$ and semi-major axis 3 times smaller than that of S2.

As is evident from Figures 8 - 10, one can obtain estimates of the
$r_c$, $\alpha$ and $\lambda _{BH}$, provided that three stars
have been observed to sufficient accuracy. Assume that we have
adequate accuracy of observation to see periastron shifts of
10$^{-2.5}$ mas, which is the value required to see the
relativistic periastron shift. To what accuracy have we limited
the cluster parameters? To determine this, we could just vary
$\lambda_{BH}$ for a given $r_c$. The effect of this change would
be less than the effect of changing $r_c$ {\it and}
$\lambda_{BH}$. As such, if we want to know how accurately the
cluster parameters are determined, we need to calculate the
maximum change in $r_c$ {\it along with} the change in
$\lambda_{BH}$, as allowed by the Brownian motion limit. By
varying $\lambda_{BH}$ by 10$^{-2}$ and $r_c$ maximally we find
that we get the required accuracy, for evaporation rates from 1 to
10 Gyr and Brownian motion of 1.3 to 2.0 km s$^{-1}$. With this
accuracy we should also be able to separate out the classical
periastron shift of the stellar cluster and the relativistic
effect of a maximal (and even slightly less than maximal) Kerr
black hole. With better accuracy we should be able to get an
estimate, or at least an upper bound, for the black hole spin as
well. The question now is, what is required to achieve this
accuracy of observation of the periastron shift of three stars?
This point is discussed in the next section.

\section{Observational requirements for determination of black
hole spin and cluster parameters}

In the near future, observations using large diameter telescopes
in combination with adaptive optics may allow us to reach the
angular resolution needed to measure the periastron shift of stars
close to the GC back hole. Consider for example an instrument with
an angular resolution $\Delta \Phi _A$ and assume that the
relative position of stars can be determined to about $1/\epsilon$
of the achieved angular resolution, i.e. the position accuracy is
$\Delta \Phi _P\simeq \Delta \Phi _A/\epsilon$. The positional
accuracy can be increased by a factor $\sqrt{N}$, if $N$ reference
stars are used. In this case, the maximum positional accuracy is
simply given by (\citealt{Rubilar01})
\begin{equation}
\Delta \phi _{P} = \frac{\Delta \Phi _P}{\sqrt{N}}~.
\end{equation}
It follows that if the periastron position of a star shifts by an
amount $\Delta \Phi _E$ (as observed from Earth), to obtain the
desired accuracy we need at least that $\Delta \phi _{P} \simeq
\Delta \Phi _E$. In this case, the minimum number of reference
stars can be determined once both the instrument angular and
positional accuracies are known, i.e.
\begin{equation}
N_{max}=\left(\frac{\Delta \Phi _P}{\Delta \Phi _E}\right)^2~.
\end{equation}

As an example, the LBT interferometer has angular resolution
$\Delta \Phi _A\simeq 30$ mas and the relative position of stars
is conservatively estimated to be about 1/30 of that value
(\citealt{Rubilar01}). Therefore, to measure the periastron shift
with adequate accuracy to see the least shift for the cluster
parameters allowed, and thereby detect the relativistic shift, we
need $N$ to be $\simeq (6.6\times10^{-1}/2\times10^{-3})^2$ or
10$^5$ reference stars, which automatically provides the accuracy
required to see the maximal Kerr (spin) effect. For PRIMA (see
http://www.eso.org/projects/vlti/instru/prima/index\_prima.html)
the relative positional accuracy is planned to be $\simeq
10~\mu$as. As such, we only need a single reference star.


\section{Concluding remarks}

We have used the fact that the stellar cluster close to the
central black hole seems spherically symmetric to limit the
Brownian motion of Sgr A$^*$ to be the observed proper motion. We
have taken the stronger (1$\sigma$) limit of 1.3 km $\rm{s}^{-1}$
and the weaker (2$\sigma$) limit of 2.0 km $\rm{s}^{-1}$ for our
calculations. We also used evaporation times of 1 to 10 Gyr for
the cluster, appropriately modified to incorporate the
gravitational well due to the black hole, to put further
constraints on the cluster mass. The results of our calculations
show that the stellar periastron shifts due to the cluster, even
limited to the extent considered, may totally swamp not only the
Kerr (spin) effect but also the Schwarzschild effect. However, the
discussion focused on the observations for a single star, S2. By
modelling  the star cluster density profile with a Plummer low,
the periastron shift contribution due to the stellar cluster
depends on three parameters: the central density $\rho _0$ (or
equivalently $\lambda _{BH}$), the core radius $r_c$, and the
power-law index $\alpha$. Consequently, with observations of three
stars we should be able to determine the cluster parameters
adequately. \footnote{Note that, wether we would have known that
the star cluster follows a Bahcall-Wolf profile, by measuring the
periastron advance of only one star we may be able to calculate
the only parameter: $\lambda_{BH}$ (from Fig. 8).} We have
addressed the question of what is required to obtain the desired
accuracy for observing the relativistic effect. It turns out that
we need about 10$^5$ reference stars with the LBT interferometer.
With the accuracy expected of PRIMA, it should be enough to use
only one reference star.

\acknowledgements

This work has been partially supported by MIUR (Programmi di
Ricerca Scientifica di Rilevante Interesse Nazionale (PRIN04) -
prot. 2004020323$\_$004). We would like to thank an unknown
referee for very useful suggestions and comments that have
improved our paper. Two of us (AQ and AFZ) would like to thank the
Department of Physics of University of Lecce and {\it INFN}
(Italy) where this work has been initiated. FDP and AAN would like
to thank the $30^{th}$ International Nathiagali Summer College
(Pakistan), where the original version of this work has been
completed. AFZ is also grateful to the National Natural Science
Foundation of China (NNSFC) (Grant \# 10233050) and National Key
Basic Research Foundation (Grant \# TG 2000078404) for a partial
financial support of the work.



\begin{thebibliography}{}

\bibitem[\protect\citeauthoryear{Bahcall \& Wolf}{1977}]{bw}
Bachall J.N. and Wolf R.A. 1977, \apj, 216, 883

\bibitem[\protect\citeauthoryear{Bini et al.}{2005}]{bini2005}
Bini D. et al. 2005, Gen. Rel. Grav., 37, 1263

\bibitem[\protect\citeauthoryear{Binney \& Tremaine}{1987}]{binneytremaine}
Binney J. and Tremaine S., {\it Galactic Dynamics}, Princeton
University Press, Princeton, New Jersey, 1987

\bibitem[\protect\citeauthoryear{Boyer \& Price}{1965}]{boyerprice}
Boyer R.H. and Price T.G. 1965, Proc. Camb. Phil. Soc. 61, 531

\bibitem[\protect\citeauthoryear{Chatterjee et al.}{2002}]{chatterje}
Chatterjee P., Hernquist L. and Loeb A. 2002, \apj, 572, 371

\bibitem[\protect\citeauthoryear{Delplancke et al.}{2003}]{Delplancke03}
Delplancke F. et al.  2003, \aap, 286, 99

\bibitem[\protect\citeauthoryear{De Paolis et al.}{2003}]{depaolis1}
De Paolis F. et al. 2003, \aap, 409, 809

\bibitem[\protect\citeauthoryear{De Paolis et al.}{2004}]{depaolis2}
De Paolis F. et al. 2004, \aap, 415, 1

\bibitem[\protect\citeauthoryear{De Paolis et al.}{2005}]{depaolis4}
De Paolis F. et al. 2005, in {\it Proc. Eleventh Regional Conf. on
Math. Phys.} eds. Rahvar S, Sadooghi N, and Shojai F, World
Scientific


\bibitem[\protect\citeauthoryear{Fabian}{2005}]{Fabian04}
   Fabian A.C. 2005, Ap\&SS, 300, 97

\bibitem[\protect\citeauthoryear{Fabian et al.}{2000}]{Fabi00}
   Fabian A.C. et al. 2000,  \pasp, 112, 1145

\bibitem[\protect\citeauthoryear{Fabian et al.}{1995}]{fabian1}
   Fabian A.C. et al. 1995, \mnras, 277, L11

\bibitem[\protect\citeauthoryear{Fragile \& Mathews}{2000}]{Fragile00}
Fragile P.C. \& Mathews G.J.  2000, \apj, 542, 328

\bibitem[\protect\citeauthoryear{Genzel et al.}{2003 a}]{Genzel03}
Genzel R. et al. 2003 a, Nature, 425, 934

\bibitem[\protect\citeauthoryear{Genzel et al.}{2003 b}]{Genzel03b}
Genzel R. et al.  2003 b, ApJ, 594, 812

\bibitem[\protect\citeauthoryear{Ghez et al.}{2003}]{Ghez03}
Ghez A.M. et al. 2003, \apjl, 586, L127

\bibitem[\protect\citeauthoryear{Ghez et al.}{2004}]{Ghez04}
Ghez A.M., et al. 2004, \apjl, 601, L159

\bibitem[\protect\citeauthoryear{Ghez et al.}{2005}]{Ghez05}
Ghez A.M., et al. 2005, \apj, 620, 744

\bibitem[\protect\citeauthoryear{Jaroszynski}{1998}]{Jaroszynski98}
Jaroszynski M.  1998, Acta Astron., 48, 653

\bibitem[\protect\citeauthoryear{Jaroszynski}{1999}]{Jaroszynski99}
Jaroszynski M.   1999, \apj, 521, 591

\bibitem[\protect\citeauthoryear{Jaroszynski}{2000}]{Jaroszynski00}
Jaroszynski M.   2000, Acta Astron., 50, 67

\bibitem[\protect\citeauthoryear{Mouawad et al.}{2005}]{moawad}
Mouawad N. et al.  2005, Astron. Nachr., 326, 83

\bibitem[\protect\citeauthoryear{Quirrenbach}{2003}]{Quirrenbach03}
Quirrenbach A.   2003, Ap\& SS, 286, 277

\bibitem[\protect\citeauthoryear{rauch}{1996}]{rauch}
Rauch  K.P. \& Tremaine S. 1996, NewA 1, 149

\bibitem[\protect\citeauthoryear{Reid and Bruthaler}{2004}]{reid2004}
Reid  M.J. and Brunthaler A. 2004, \apj, 616, 872

\bibitem[\protect\citeauthoryear{Reid et al.}{1999}]{reid1999}
Reid  M.J. et al. 1999, \apj, 524, 816

\bibitem[\protect\citeauthoryear{R\"{o}ttgering et al.}{2003}]{Rottgering03}
R\"{o}ttgering H.J.A. et al. 2003, astro-ph/0308538

\bibitem[\protect\citeauthoryear{Rubilar \& Eckart}{2001}]{Rubilar01}
Rubilar G.F. \& Eckart A.  2001, \aap, 374, 95

\bibitem[\protect\citeauthoryear{Sch\"odel et
al.}{2003}]{Schoedel03} Sch\"odel R. et al.  2003, \apj, 596, 1015

\bibitem[\protect\citeauthoryear{Shen et
al.}{2005}]{shen} Shen Z.-Q. et al.  2005, Nature, 438, 62


\bibitem[\protect\citeauthoryear{Smart}{1977}]{smart}
Smart W.M. 1977, {\it Textbook on Spherical Astronomy}, Cambridge
University Press

\bibitem[\protect\citeauthoryear{Tanaka et al.}{1995}]{tanaka1}
Tanaka  Y. et al. 1995,  Nature,  375, 659

\bibitem[\protect\citeauthoryear{Weinberg, Miloslavljevi\'{c} \& Ghez}{2005}]{Weinberg05}
Weinberg N.N., Miloslavljevi\'{c} M.\& Ghez A.M.  2005, \apj, 622,
878

\bibitem[\protect\citeauthoryear{Weinberg}{1972}]{Weinberg72}
Weinberg S. 1972, Gravitation and Cosmology: Principles and
Applications of the General Theory of Relativity, Wiley, New York

\bibitem[\protect\citeauthoryear{Zakharov et al.}{2003a}]{ZKLR02}
   Zakharov A.F.et al. 2003a, \mnras, 342, 1325

\bibitem[\protect\citeauthoryear{Zakharov et al.}{2005a}]{ZNDI_04}
Zakharov A.F.et al. 2005a, New Astronomy, 10, 479

\bibitem[\protect\citeauthoryear{Zakharov et al.}{2005b}]{depaolis3}
Zakharov A.F. et al. 2005b, \aap, 442, 795

\bibitem[\protect\citeauthoryear{Zakharov \& Repin }{2003b}]{zak_rep03_aa}
   Zakharov A.F. \& Repin S.V. 2003b, \aap, 406, 7

\bibitem[\protect\citeauthoryear{Zakharov \& Repin }{2003c}]{Zak_rep03_AR}
   Zakharov A.F. \& Repin S.V. 2003c, Astron. Rep., 47, 733

\bibitem[\protect\citeauthoryear{Zakharov \& Repin }{2004}]{ZR_ASR04}
 Zakharov A.F. \& Repin S.V. 2004, Adv. Space Res.,  34, 1837

\end{thebibliography}
\end{document}